\documentclass{aa}                                          
\pdfoutput=1																							  

\usepackage[english]{babel}   
\usepackage[utf8]{inputenc}   
\usepackage{amsmath,amssymb,amsfonts}    
\usepackage[version=3]{mhchem}           
\usepackage{siunitx}                     
\usepackage{multirow}         
\usepackage{graphicx}         
\usepackage[varg]{txfonts}              

\usepackage{natbib}                     
\bibpunct{(}{)}{;}{a}{}{,}

\everymath{\displaystyle}

\usepackage{cleveref} 

\begin{document}


\title{Evidence against a strong thermal inversion  \\
       in HD 209458 b from high-dispersion spectroscopy \thanks{Based on observations collected at the ESO Very Large Telescope (Program 186.C-0289)}}
\author{Henriette Schwarz\inst{1}
\and Matteo Brogi\inst{1,4}
\and Remco de Kok\inst{1,2}
\and Jayne Birkby\inst{1,3,5}
\and Ignas Snellen\inst{1}}
\institute{Leiden Observatory, Leiden University, P.O. Box 9513, 2300 RA Leiden, The Netherlands
\and SRON Netherlands Institute for Space Research, Sorbonnelaan 2, 3584 CA Utrecht, The Netherlands
\and Harvard-Smithsonian Center for Astrophysics, 60 Garden Street, Cambridge MA, 02138, USA
\and NASA Hubble Fellow
\and NASA Sagan Fellow}
\date{}

\abstract
{Broadband secondary-eclipse measurements of strongly irradiated hot Jupiters have indicated the existence of atmospheric thermal inversions, but their presence is difficult to determine from broadband measurements because of degeneracies between molecular abundances and temperature structure. Furthermore, the primary mechanisms that drive the inversion layers in hot-Jupiter atmospheres are unknown. This question cannot be answered without reliable identification of thermal inversions.
}
{We apply high-resolution ($R=\num{100000}$) infrared spectroscopy to probe the temperature-pressure profile of HD 209458 b. This bright, transiting hot-Jupiter has long been considered the gold standard for a hot Jupiter with an inversion layer, but this has been challenged in recent publications.
}
{We observed the thermal dayside emission of HD 209458 b with the CRyogenic Infra-Red Echelle Spectrograph (CRIRES) on the Very Large Telescope during three nights, targeting the carbon monoxide band at \SI{2.3}{\micro\m}. Thermal inversions give rise to emission features, which means that detecting emission lines in the planetary spectrum, as opposed to absorption lines, would be direct evidence of a region in which the temperature
increases with altitude. 
}
{We do not detect any significant absorption or emission of \ce{CO} in the dayside spectrum of HD 209458 b, although cross-correlation with template spectra either with \ce{CO} absorption lines or with weak emission at the core of the lines show a low-significance correlation signal with $\textrm{a signal-to-noise ratio of} \sim$ 3 -- 3.5. Models with strong \ce{CO} emission lines show a weak anti-correlation with similar or lower significance levels. Furthermore, we found no evidence of absorption or emission from \ce{H2O} at these wavelengths. 
}
{The non-detection of \ce{CO} in the dayside spectrum of HD 209458 b is interesting in light of a previous \ce{CO} detection in the transmission spectrum. That there is no signal indicates that HD 209458 b either has a nearly isothermal atmosphere or that the signal is heavily muted. Assuming a clear atmosphere, we can rule out a full-disc dayside inversion layer in the pressure range \SI{1}{\bar} to \SI{1}{mbar}.
}


\titlerunning{Evidence against a strong thermal inversion in HD 209458 b}
\authorrunning{Schwarz et al.}

\maketitle


\section{Introduction}

\subsection{Hot-Jupiter atmospheres}

Hot Jupiters form an ideal starting point for exoplanet atmospheric research. While they experience high levels of stellar irradiation, their atmospheric scale heights are relatively large, and their high temperatures result in high planet-to-star contrast ratios, especially in the infrared. Furthermore, since hot Jupiters have short orbital periods (1-5 days), they have a high probability of transiting their host star, and transits and eclipses occur on conveniently short timescales. The accessibility of hot Jupiters have made them excellent test objects for developing techniques for atmospheric characterisation - which are applicable to the next generation of telescopes and instruments and hence also to smaller planets \citep[e.g.][]{Schneider1994, Webb2001, Seager2010, Snellen2013a}. The composition and structure of individual hot-Jupiter atmospheres are themselves of great interest to better understand the complex physical processes involved (including strong stellar irradiation, atmospheric circulation, and photochemical processes), and because they may hold clues to the planetary formation history \citep[e.g.][]{Oberg2011, Madhusudhan2014}.

Atmospheric characterisation of hot Jupiters has so far mainly focused on transiting planets because they offer unique observational possibilities. When a planet transits its host star, part of the stellar light is filtered through the planetary atmosphere, leaving an absorption fingerprint in the observed light \citep{Seager2000, Brown2001, Charbonneau2002}. At the opposite side of the orbit, the star occults the planet, providing an opportunity to isolate the planetary contribution to the total flux. The difference between observations of the combined light of the star and planet in and out of secondary eclipse can be used to determine the planet's dayside spectrum \citep{Charbonneau2005, Deming2005b}. In addition, flux variations during the entire orbit can be measured as a function of phase, gaining information on the day- and nightside contributions \citep[e.g.][]{Knutson2007b, Snellen2009}.

Hot Jupiters are assumed to have atmospheres that are dominated
by molecular hydrogen, but the spectrum of a given planet will contain several spectroscopically active trace gases of carbon- and oxygen-bearing molecules. At high temperatures and observed in the infrared, the most abundant of these molecules are expected to be \ce{CO} and \ce{H2O}, followed by \ce{CH4}. Other significant molecules are possibly \ce{CO2}, \ce{C2H2}, and \ce{HCN} \citep{Madhusudhan2012, Moses2013}.

\subsection{Thermal inversion layers}

The upper atmospheres of hot Jupiters are thought to be radiation dominated, with the radiation efficiency depending critically on the emission and absorption of the specific molecules that are present. Generally, the radiation from the host star is expected to deeply penetrate the atmosphere and heat it from below, which
results in  temperatures that decrease with increasing altitude \citep{Guillot2010}. However, strong optical or UV absorbers in the upper layers of the atmosphere can cause thermal inversion - a layer or region in the atmosphere in which the temperature instead increases with altitude \citep{dePater2010book}. Thermal inversions are common among the solar system planets. Jupiter, along with the other giant planets, has thermal inversions caused by \ce{CH4}-induced hazes, while the inversion in the terrestrial stratosphere is caused by \ce{O3} \citep[e.g.][and references therein]{Seager2010book}.

Thermal inversions may be commonly present in hot-Jupiter atmospheres as well. Secondary-eclipse measurements with the Spitzer Space Telescope have made it possible to probe the vertical temperature structure of bright transiting hot Jupiters. By measuring the relative depths of a secondary eclipse in multiple band-passes, a low-resolution thermal spectrum of the exoplanet can be constructed that can then be compared with theoretical spectra. There have been several reports of inversion layers in hot Jupiters from Spitzer observations (HD 209458b, \citealt{Burrows2007, Knutson2008}; XO-1b, \citealt{Machalek2008}; XO-2b, \citealt{Machalek2009}; TrES-4, \citealt{Knutson2009}; TrES-2, \citealt{ODonovan2010}; Hat-P-7b, \citealt{Christiansen2010}), while other planets show no sign of an inversion \citep[e.g. HD 189733b,][]{Charbonneau2008}. The inference of a thermal inversion is in most cases based upon an excess of flux in the \SI{4.5}{\micro\m} and \SI{5.8}{\micro\m} Spitzer band-passes, which has been interpreted as due to water emission features. However, many of these claims are based upon the comparison of only a few inverted and non-inverted atmospheric models, which may not adequately map degeneracies between atmospheric parameters. In contrast, a systematic retrieval analysis of secondary-eclipse spectra of nine hot Jupiters (\citealt{Line2014}) found little evidence for thermal inversions over a wide range of effective temperatures (with the exception of HD209458b).

The nature of the possible responsible absorbers is as yet unknown in this high-temperature regime. It has been proposed that highly irradiated planets are warm enough for significant amounts of gaseous \ce{TiO} and \ce{VO} to exist in the upper atmosphere, providing the necessary opacity to generate a temperature inversion \citep{Hubeny2003, Fortney2006, Fortney2008, Burrows2008a}. This advocates a correlation between high irradiation and existence of inversion layers. No observational evidence for \ce{TiO} and \ce{VO} absorption in hot Jupiter atmospheres has been found,
however \citep{Huitson2013, Sing2013, Gibson2013, Bento2014, Hoeijmakers2014}, although \citet{Desert2008} suggested \ce{TiO} and \ce{VO} molecules as possible candidates to explain observed absorption broadband features in the optical transmission spectrum of HD 209458 b.

The \ce{TiO}/\ce{VO} hypothesis has since been challenged by \citet{Spiegel2009}, who argued that even a ten times supersolar abundance of \ce{VO} is insufficient to drive a thermal inversion  and that the high molecular weight of \ce{TiO} requires substantial vertical mixing to retain a high abundance of \ce{TiO} at low pressures. Apart from a deep vertical cold trap, \ce{TiO} can be depleted by a nightside cold trap if it condenses to particles larger than a few microns \citep{Parmentier2013b}. Furthermore, \citet{Madhusudhan2011} found that for carbon-rich (\ce{C}/\ce{O} > 1) hot Jupiters, most of the oxygen is bound up in \ce{CO}, and as a result, the abundances of \ce{TiO} and \ce{VO} are too low to cause a thermal inversion. Alternatively, \citet{Knutson2010} found a preference for planets thought to host an inversion to orbit around chromospherically quiet stars and vice versa, and they theorised that increased UV levels for active stars were destroying the unknown compounds responsible for inversion layers. The inversion-causing absorbers could also be the result of non-equilibrium processes. \citet{Zahnle2009} propounded the photochemically produced sulfur compounds, \ce{S2} and {HS}, as efficient UV and optical absorbers at high temperatures.

Comparative studies of hot Jupiters with and without temperature inversions are key to explaining the physical causes of inversion layers. Although such studies have been attempted \citep{Fortney2008, Burrows2008a, Knutson2010}, this is not truly possible until a reliable method is available for measuring thermal inversions. Broadband photometry does not resolve the molecular spectral bands, which means that emission lines cannot be directly detected, and the inference of a thermal inversion must therefore rely heavily on models and their inherent assumptions. The Spitzer inferences of inversion layers tend to rely on assumptions about the presence of water, and typically, solar composition is assumed. \citet{Madhusudhan2010} explored a wider paramater space and found evidence for degeneracies between the temperature pressure profile and the molecular abundances.

\subsection{High-resolution spectroscopy}

Ground-based high-dispersion spectroscopy in the near-infrared has recently become successful in characterising hot Jupiters. By resolving molecular features into individual lines, the lines can be reliably identified through line matching \citep{Brown2002, Deming2005a, Barnes2007}. \citet{Snellen2010} employed the Cryogenic Infra-Red Echelle Spectrograph \citep[CRIRES,][]{Kaeufl2004} at the Very Large Telescope (VLT) to observe the transmission spectra of HD 209458 b with a resolution of $R=\num{100000}$, and detected \ce{CO} at \SI{2.3}{\micro\m}. The technique makes use of the changing radial velocity of the planet during the observations to separate the planetary lines from the telluric and stellar lines. 

The target planet can be observed favourably either during a transit (transmission spectroscopy) or at phases shortly before or after superior conjunction (dayside spectroscopy). Since a secondary eclipse is not required to separate the contamination from the parent star, dayside spectroscopy can also be applied to non-transiting planets. $\tau$ Bo{\"o}tis b has been observed at \SI{2.3}{\micro\m} with high resolution by \citet{Brogi2012} and \citet{Rodler2012}, who detected absorption of \ce{CO}, and the radial velocity measurements led to the first mass determination of a non-transiting hot Jupiter. The dayside spectrum of 51 Peg b revealed absorption from \ce{CO} and possibly water \citep{Brogi2013}, but the signal was not detected in one of the three nights of observations, and therefore more data are necessary to confirm the result. \ce{CO} absorption was detected in \SI{2.3}{\micro\m} dayside observations of HD 189733 b \citep{deKok2013}. \ce{H2O} absorption was detected in \SI{3.2}{\micro\m} dayside observations of the same planet \citep{Birkby2013}, demonstrating the ability of high-resolution spectroscopy to also detect a more complex molecule in a wavelength regime more heavily contaminated by telluric lines. Recently, a combined signal from \ce{CO} and \ce{H2O} absorption was detected in \SI{2.3}{\micro\m} dayside observations of HD 179949 b \citep{Brogi2014}, and \citet{Lockwood2014} detected \ce{H2O} in the dayside spectrum of $\tau$ Bo{\"o}tis b at L-band using NIRSPEC at Keck. 

Dayside high-dispersion infrared spectroscopy has the potential to provide evidence for an inversion layer in a hot Jupiter. The molecular bands are resolved into individual lines, so that the average shape of the lines and the average contrast with respect to the stellar continuum can be measured, or at least constrained. The dayside spectrum of a hot Jupiter is sensitive to the presence of thermal inversions because the line-of-sight to a high degree coincides with the vertical direction in the planetary atmosphere. The sign of the vertical temperature gradient determines whether the thermal emission from the molecules in the upper atmosphere is seen as emission or absorption relative to the continuum part of the spectrum, emitted deep in the atmosphere (between $1$ and $0.1$ bar). The sign of the measured line contrast thus allows distinguishing between emission and absorption lines, and the shape of the lines can help further constrain the temperature-pressure profile.

\subsection{HD 209458 b}
\label{subsec:HD209b}

At this point, all published high-dispersion detections of molecules have detected lines in absorption, but if instead emission lines were to be detected, this would prove the existence of a thermal inversion. We therefore apply here the high-resolution technique to the dayside of the hot Jupiter HD 209458 b. The planet was the first exoplanet known to transit its parent star \citep{Charbonneau2000}, and today it is one of the best-studied hot Jupiters.

In the context of inversion layers, HD 209458 b is of particular interest since it was the first exoplanet reported to have a thermal inversion \citep{Burrows2007, Knutson2008}, and until recently, it was often proclaimed to be the benchmark for a hot Jupiter with a thermal inversion in its upper atmosphere. This was based on secondary-eclipse observations at \num{3.6}, \num{4.5}, \num{5.8} and \SI{8.0}{\micro\m} using the Spitzer Infrared Array Camera \citep[IRAC][]{Knutson2008}. The \num{4.5} and \SI{5.8}{\micro\m} bandpasses showed an excess of flux relative to the \SI{3.6}{\micro\m} bandpass, in contrast to traditional models that expected a trough at these wavelengths due to water absorption features. The excess flux was best explained with a low-pressure inversion layer producing water emission features. The claim of a thermal inversion in HD 209458 b was supported by extensive modelling by \citet{Madhusudhan2009, Madhusudhan2010} and \citet{Line2014}. However, the existence of a thermal inversion in the atmosphere of HD 209458 b has recently been challenged. \citet{Zellem2014} measured the full orbit \SI{4.5}{\micro\m} phase curve with Spitzer IRAC using the now standard staring-mode observations and pixel-mapping techniques. They found the \SI{4.5}{\micro\m} secondary-eclipse depth to be 35\% shallower than previously measured by \citet{Knutson2008}, who did not use staring mode, nor intra-pixel sensitivity maps. This revision does not rule out an inversion layer, but is consistent with models both with and without a thermal inversion, or with a blackbody. \citet{Diamond-Lowe2014} have performed a self-consistent analysis of all available Spitzer/IRAC secondary-eclipse data of HD 209458 b. This includes both a reanalysis of the eclipses in the four IRAC bandpasses published by \citet{Knutson2008} and five previously unpublished eclipses in staring mode, one in 2007 at \SI{8.0}{\micro\m}, two in 2010 at \SI{4.5}{\micro\m,} and two in 2011 at \SI{3.6}{\micro\m}. They found no evidence for a thermal inversion in the atmosphere of HD 209458 b. Their best eclipse depths are well fitted by a model where the temperature decreases between pressure levels of 1 and 0.01 bars.

\subsection{Re-evaluation of the HD209458b CO abundance from \citet{Snellen2010}}

Carbon monoxide absorption has previously been observed with CRIRES at the VLT during a transit of HD 209458 b \citep{Snellen2010}, with a relative line strength at the level of \num{1} -- \num{1.5e-3} with respect to the host star spectrum. Although a constraint on the \ce{CO} volume-mixing ratio was presented, in hindsight it became clear that the underlying analysis did not include a thorough investigation of the parameter space. Furthermore, an erroneous conversion factor in the computation of the pressure broadening of the absorption lines resulted in a too high derived abundance. We note that the conversion error was only present in \citet{Snellen2010} and not in any of our subsequent work. We since performed a re-analysis of the constraints to the CO volume mixing ratio that can be inferred from the high-dispersion transit observations. A large grid of models was constructed, assuming a range of temperature-pressure profiles, abundances of a range of molecules, and pressure levels of the continuum. This re-analysis indicates that the observed relative line strength corresponds to a \ce{CO} volume-mixing ratio in the range \num{e-5} -- \num{e-4}, with an additional uncertainty of a factor of $\sim$2 if uncertainties in the high-altitude atmospheric temperature at the terminator region are included. However, if clouds or hazes
contribute to the continuum extinction, the corresponding \ce{CO} abundance could in turn be significantly higher.

\subsection{Outline} 

In this paper we present three nights of CRIRES observations of HD 209458 at \SI{2.3}{\micro\m}, targeting \ce{CO} in the planet's dayside, allowing us to probe the vertical temperature structure of the atmosphere. Section \ref{sec:observations} details our observations of HD 209458 b, and Sect. \ref{sec:data-analysis} describes the data analysis steps from the raw spectra to removal of telluric contamination. In Sect. \ref{sec:search-for-the-planet-signal} the cross-correlation analysis we used to extract the potential planetary signal is explained, and the results are presented in Sect. \ref{sec:results}. We discuss these in Sect. \ref{sec:discussion}, and conclude in Sect. \ref{sec:conclusion}.

\begin{table}[ht]
\caption{Details of the observations, showing the observing date for the beginning of the night in local time, the use of nodding, the planetary phase, the total observing time, the number of observed spectra, range in airmass, water vapour, detector integration time (DIT), number of integrations per nod or observation (NDIT), and detector counts.}                   
\label{table:observations}                                                                      
\centering       
\begin{tabular}{r c c c}                
\hline\hline                            
\\
           & 2011-08-04 & 2011-09-05 & 2011-09-12\\
\hline                            
\\
Nodding    &   yes      & yes        & no \\
Orbital Phase & 0.51--0.57 &0.55--0.62 &0.54--0.61 \\
 $\textrm{t}_\textrm{obs}$ [h] & 6.07 & 5.53 & 6.02\\
 $\textrm{N}_\textrm{obs}$ & 118 & 108 & 122\\
Airmass     & 2.16--1.38 & 2.27--1.38 & 2.20--1.38\\
DIT [s]   &  30        &  30        & 50 \\
NDIT        &  5         &   5        &  3\\
 Cnts [\#/pix/s]& 40--65 & 50--70&50--90\\

\hline                            
\end{tabular}
\end{table}

\section{Observations}
\label{sec:observations}

We observed HD 209458 ($K = 6.31$ mag) for a total of $17.5$ hours during three nights in August and September 2011 as part of the large ESO program 186.C-0289, which was designed to study the brightest transiting  and non-transiting hot-Jupiter systems visible from Cerro Paranal. The observations were obtained with the CRIRES instrument, which is mounted at the Nasmyth A focus of VLT Antu. We observed at the standard wavelength settings for the reference wavelength \SI{2.3252}{\micro\m}, covering the ro-vibrational (2,0) R-branch of carbon monoxide. The CRIRES instrument has four Aladdin III InSb detectors, each of the size 1024x512 pixels with gaps of about \num{280} pixels between the individual detectors. 

We used a \ang{;;0.2} slit in combination with the Multi Application Curvature Adapative Optics system \citep[MACAO,][]{Arsenault2003} to achieve the highest possible resolution of $R \simeq \num{100000}$. During each of the three nights, the target was observed continuously for 5--6 hours at planetary phases shortly after a secondary eclipse, when a significant portion of the dayside of the planet was visible. Additionally, a standard set of calibration frames was obtained during daytime and twilight. 

The first two nights were observed in nodding mode, where the telescope is nodded along the slit by \ang{;;10} in an ABBA pattern, providing an easy method for background subtraction. The third night was observed without nodding in an effort to  enhance stability and reduce overheads. Details of the observations are given in \Cref{table:observations}.

\section{Data analysis}
\label{sec:data-analysis}

The data in this work have been analysed in a similar way as in \citet{Snellen2010} and \citet{Brogi2012, Brogi2013}, but all analyses were performed with new purpose-built Python scripts.

\subsection{Extracting the one-dimensional spectra}
\label{subsec:extraction-of-the-1D}

The basic image processing was performed with the CRIRES pipeline version 2.1.3 and the corresponding version 3.9.0 of ESOREX. The images were dark-subtracted and flatfielded and were corrected for bad pixels and non-linearity effects. For the first two nights, the CRIRES pipeline was also used to combine the images in AB nodding pairs, thus performing a background subtraction, and to extract the 1D spectra. This resulted in a single extracted spectrum for every two observations. These spectra were extracted with the optimal extraction technique \citep{Horne1986}. The spectra of the final night were observed without nodding, for which the CRIRES pipeline was ill-suited. Instead, we used the IRAF data analysis package {\sl apall}, both for the background subtraction and the optimal extraction of the spectra.

Detectors 1 and 4 are both affected by the odd-even effect, a non-linear change in gain between odd and even columns along the spectral direction\footnote{http://www.eso.org/sci/facilities/paranal/instruments/crires/doc/VLT-MAN-ESO-14500-3486\_v93.pdf}. This is the result of the reading direction being perpendicular to the spectral dispersion for these two detectors. In particular, detector 4 showed residual odd-even imprints after non-linearity correction. Furthermore, detector 1 has very few expected carbon monoxide lines. For these reasons, we chose to leave out detectors 1 and 4 from further analysis, which now covers the wavelength range \SIrange[range-phrase = --]{2.3038}{2.3311}{\micro\m}.

\begin{figure}
\centering
        \includegraphics[width=0.9\hsize]{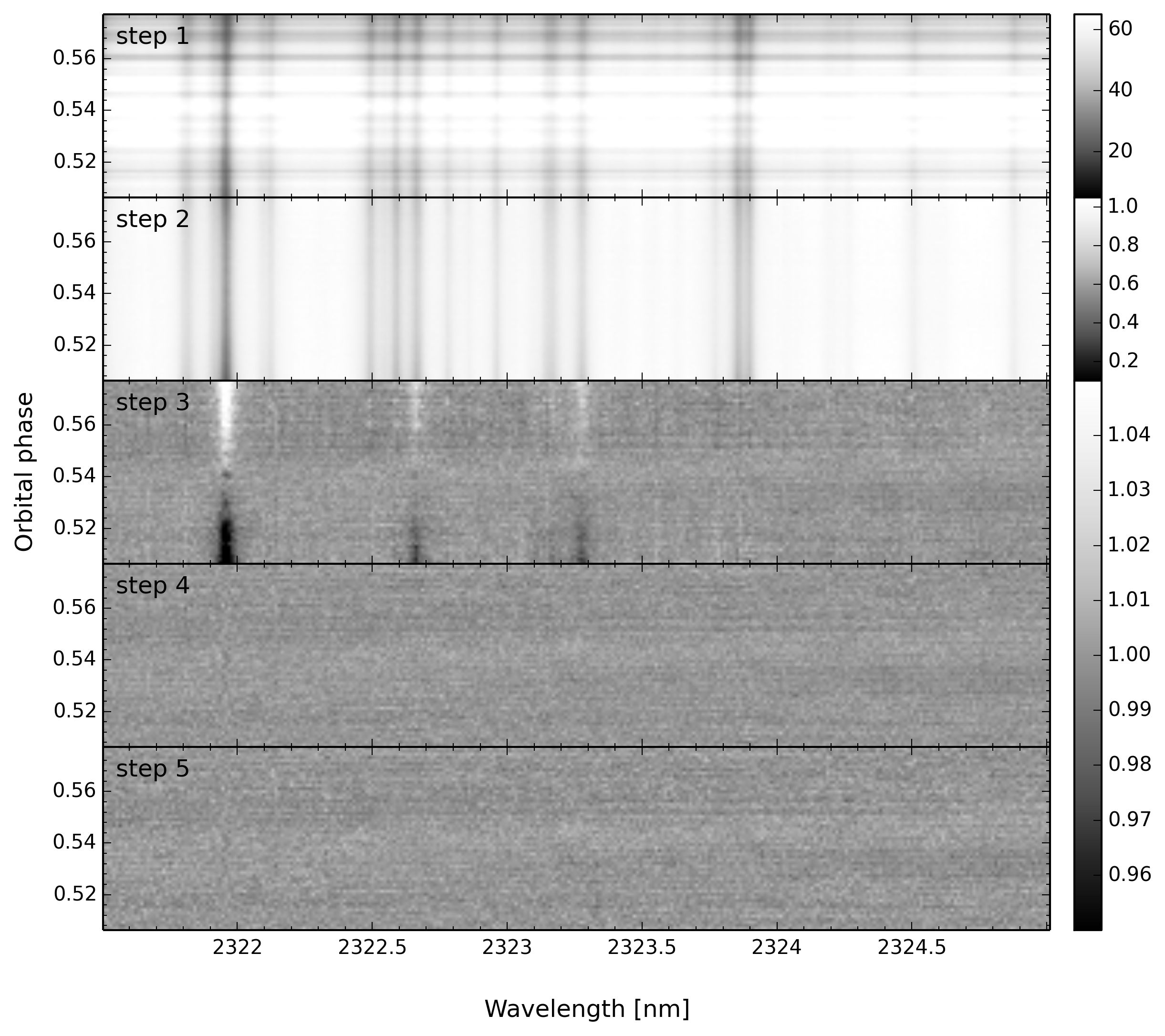}
\caption{Overview of the different steps in the data analysis, showing a small fraction of the third array from the first night of observations. From top to bottom, step 1 shows the extracted spectra after bad-pixel correction, step 2 after spectral alignment and normalization, step 3 after first-order airmass correction, step 4 after second-order airmass correction, and step 5 after normalization of each column with the variance.}
\label{fig:waterfall}
\end{figure}

\subsection{Bad-pixel correction and wavelength calibration}
\label{subsec:bad-pixel-wavelength-calib}

The extracted spectra were treated separately per night and per detector. The spectra were arranged into two-dimensional matrices with wavelength along the x-axis and time along the y-axis. The bad-pixel correction included in the CRIRES pipeline is insufficient, and we therefore manually identified bad pixels, columns, and regions through visual inspection of the matrices with the program DS9. A few larger regions consisting of several adjacent bad columns, such as the columns closest to the edge of the CCD's and a well-known defect on the second detector, were masked during the subsequent data analysis. The masked pixels made up about 2\% of the total number of pixels. Remaining individual bad pixels and bad columns were corrected with cubic spline interpolation based on the four nearest neighbours on each side in the row. 

All spectra of a given matrix were aligned to a common wavelength grid. We determined the offsets between centroids (determined using a Gaussian fit) of deep and isolated telluric lines in the individual spectra (typically 15-20 per array) and those in the average spectrum. The centroid offsets from a given spectrum showed no trend with wavelength and a typical scatter of less than 0.1 pixel with respect to each other. Each spectrum was therefore shifted with a spline interpolation using the median centroid offset of that spectrum. Next, we matched the pixel positions of centroids of telluric lines in the average aligned spectrum with the wavelengths in a synthetic telluric transmission spectrum from ATRAN\footnote{http://atran.sofia.usra.edu/cgi-bin/atran/atran.cgi} \citep{Lord1992}. The pairs of pixel position and wavelength were fitted with a second-order polynomial to yield the wavelength solution for a given night and detector. The highest residuals to the second-order fits were of the order \SI{\pm 3e-6}{\micro\m}, which corresponds to 20\% of a pixel. Finally, each spectrum was normalised by its median value.

\subsection{Removing telluric contamination}
\label{subsec:removal-of-telluric-contamination}

The observed near-infrared spectra are dominated by absorption lines originating in the atmosphere of Earth, and a crucial datareduction step is therefore to remove this telluric contamination. The telluric lines are stationary in wavelength (but vary in strength) over the course of the observations, and therefore fall on fixed columns in the matrices. On the other hand, molecular lines arising in the planetary atmosphere are Doppler-shifted in each consecutive spectrum as the radial component of the orbital velocity of the planet changes, resulting in diagonal features across the matrices, buried in the noise. The planet lines typically move by 30 to 40 pixels during one night of observation. This means that we can manipulate the matrices column by column without significantly affecting the planetary signature.  

The variations along columns of the matrices are strongly correlated with the changing airmass over the time interval of the observations. We corrected for this with a linear least-squares fit of the logarithm of the fluxes in each column as a function of geometric airmass.

At the positions of some of the more prominent telluric lines, there are clear residuals, showing a correlation with time. These second-order effects stem from changing conditions in Earth's atmosphere over the course of the night. Two distinct patterns in time were recognizable and were found to correspond to water and methane lines. We sampled the two effects of flux with time in a few affected columns in a strong water line and a strong methane line. We then corrected the remaining matrix by expressing each column as a linear function of the sampled effects and dividing by the fit. The second-order airmass fitting can be affected by the planetary signal we are ultimately looking for. In particular, in columns (wavelengths) where the planet signal is at the start or end of the time series, the fit can be affected by this additional signal and therefore partially compensate for it. This is overall typically a 10\% effect. Therefore we masked pixels expected to contain a planetary signal when performing this step.

\begin{figure*}
\centering
        \includegraphics[width=17cm]{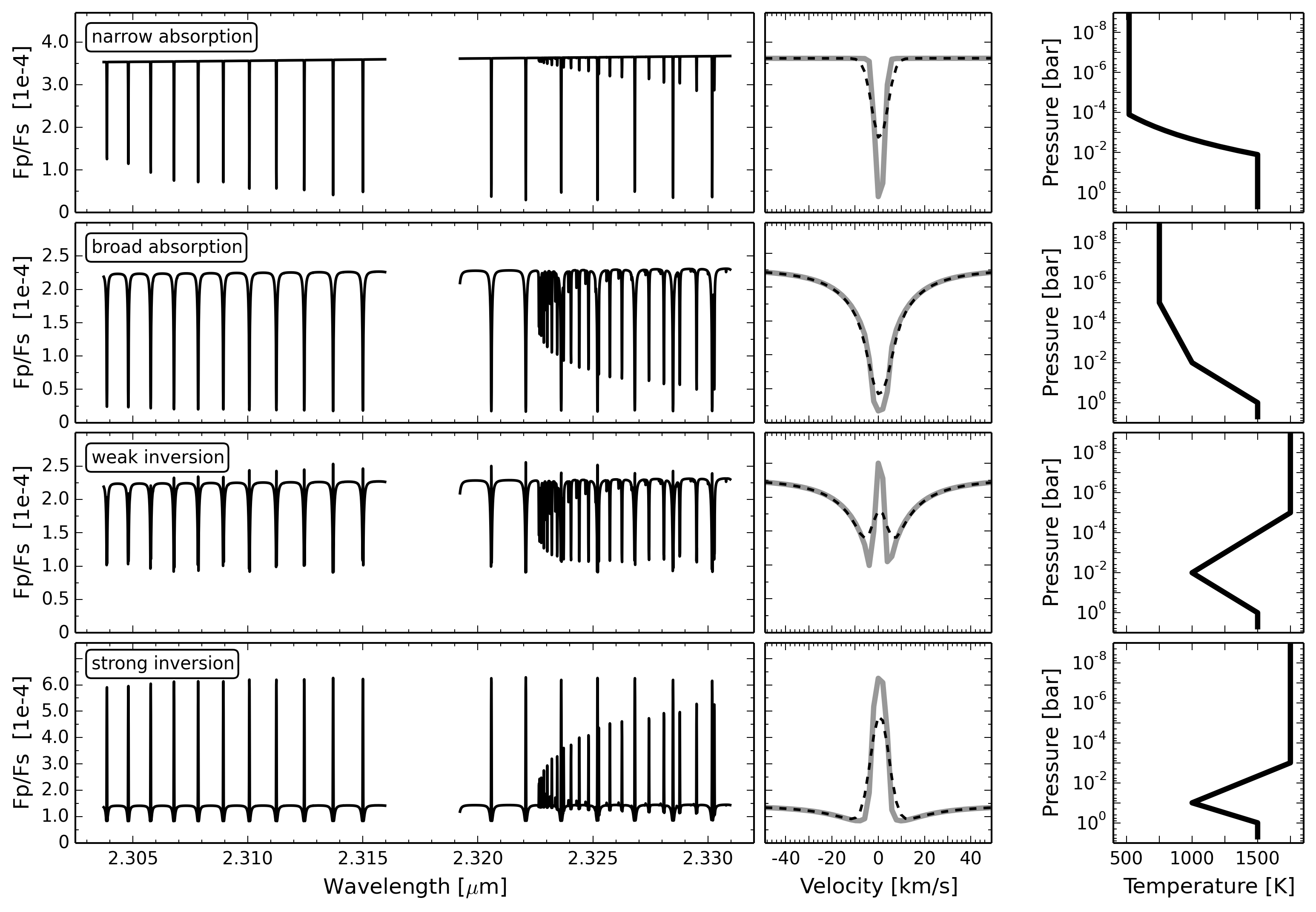}
        \caption{Examples of \ce{CO} model spectra used in the cross-correlation analysis. The model in the top row has an adiabatic lapse rate and a \ce{CO} $\textrm{VMR}$ of $\num{e-4.5}$, resulting in strong, narrow absorption lines. The following three rows are models from the T/P grid described in Sect. \ref{subsec:the-model-spectra}, all with \ce{CO} $\textrm{VMR}=\num{e-4.0}$: the first is without a thermal inversion and with relatively broad absorption lines, the second is with a weak inversion, and the last model is for a strong thermal inversion. The left column displays the models before convolution to CRIRES resolution over the wavelength coverage of the two central detectors. The middle column zooms in on the \ce{CO} line at approximately \SI{2.3205}{\micro\m}, showing the line profiles both before (solid) and after (dashed) convolution to CRIRES resolution. The right column gives the corresponding T/P profiles.}
        \label{fig:model-spectra}
\end{figure*}

Following this, a few columns that coincided with the wavelengths of the deepest telluric lines were visibly more noisy than the rest. We divided each column by the variance of itself to ensure that the noisiest columns did not dominate the cross-correlation in the next step of the data analysis. An example of the intermediate results of the different steps in the data reduction is shown in Fig. \ref{fig:waterfall}. We achieve a typical signal-to-noise ratio (S/N) of $\sim$250 per pixel per spectrum for the first two nights, and $\sim$200 per pixel per spectrum for the final night, for which we did not dither.

\section{Search for the planet signal}
\label{sec:search-for-the-planet-signal}

The individual planetary lines are buried in the noise of the residual spectra. With the chosen wavelength setting of \SI{2.3}{\micro\m}, we are targeting the (2,0) R-branch of \ce{CO}, and we expect tens of \ce{CO} lines within the observed wavelength range (see Fig. \ref{fig:model-spectra}). The signals from the individual lines are combined by cross correlating the residual spectra one at a time with purpose-built model spectra. In this section we describe the model spectra and the cross-correlation analysis used to search for a planetary signal in the observed spectra.

\subsection{Model spectra}
\label{subsec:the-model-spectra} 

The model spectra of HD 209458 b were calculated line by line using parameterised temperature-pressure profiles (T/P profiles), \ce{H2}-\ce{H2} collision-induced absorption \citep{Borysow2001, Borysow2002} and with either \ce{CO} or \ce{H2O} as a single trace gas. The \ce{CO} and \ce{H2O} data were taken from HITEMP 2010 \citep{Rothman2010}, and a Voigt line profile was employed. We used the parameter values for the Voigt profile as given by the HITEMP database. Uncertainties in the line shape parameters will mainly influence the depths of the lines, not the cross-correlation values, since the CO lines are widely separated, and the spectral shape will not be affected much. In this article, we focus on the analysis and results from cross correlating with the model spectra with \ce{CO} as the single trace gas. A short summary of the results from the \ce{H2O} models, which showed no signs of a signal, is given in Sect. \ref{subsec:water-models}. 

We performed the cross-correlation analysis for a grid of models spanning a range of T/P profiles and \ce{CO} volume-mixing ratios (VMR). The T/P profiles describe the average vertical temperature structure of the dayside of the planetary atmosphere. All the models in the grid are isothermal with a temperature of $T_1=[\SI{1500}{K}, \SI{2000}{K}]$ at pressures higher than $P_1=\SI{1}{bar}$. The temperature decreases with a constant lapse rate (i.e. the rate of temperature change with log pressure) until it reaches $T_2=\SI{1000}{K}$ at pressure $P_2=[\SI{e-1}{bar}, \SI{e-2}{bar}]$. At higher altitudes there is either an inversion layer or not, depending on the temperature $T_3=[\SI{750}{K}, \SI{1750}{K}]$ at the pressure $P_3=[\SI{e-3}{bar}, \SI{e-5}{bar}]$, above which the models again become isothermal. This amounts to eight different T/P profiles with an inversion layer and eight T/P profiles without, all of which were tested with three different volume-mixing ratios, $\textrm{VMR}=[\num{e-4.5}, \num{e-4.0}, \num{e-3.5}]$.  

The three bottom rows of Fig. \ref{fig:model-spectra} show representative examples of \ce{CO} model spectra from the T/P grid: A model atmosphere without a thermal inversion, and two with inversion layers, all with \ce{CO} $\textrm{VMR}=\num{e-4.0}$. The model without an inversion has absorption lines, while the two models with thermal inversions conversely show emission at the core of the \ce{CO} lines and absorption in the wings. In the following, we refer to the model with the strongest emission lines as the strong-inversion model and to the one with weaker emission as the weak-inversion model. The distinction between weak and strong is based on whether or not the core emission is below or above the continuum level after convolving to the resolution of the CRIRES instrument. All the models from the grid without an inversion layer have relatively broad \ce{CO} absorption lines (FWHM $\sim$ \num{10} to \SI{45}{\kilo\m\per\s}), which have a tendency to smear out the planetary signal in the cross-correlation analysis. We therefore also investigated a \ce{CO} model with more narrow absorption lines (FWHM $\sim$4 km sec$^{-1}$), illustrated in the top row of Fig. \ref{fig:model-spectra}. This model has no thermal inversion and an adiabatic lapse rate until the assumed lowest temperature of \SI{500}{K}, thus maximising the temperature gradient for the given \ce{CO} volume-mixing ratio, which in this case is \num{1e-4.5}.

\subsubsection{More inversion models}
\label{subsubsec:more-inversion-models}

\begin{figure}[ht]
\resizebox{0.9\hsize}{!}{\includegraphics{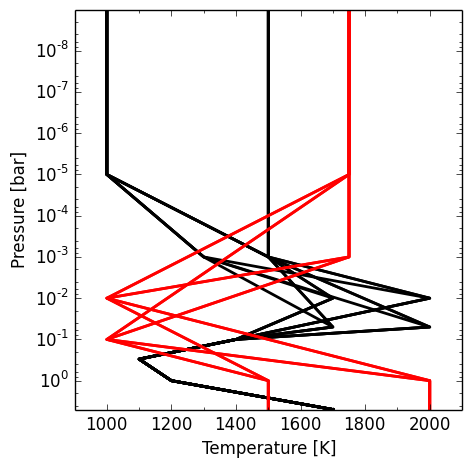}}
\caption{Temperature-pressure profiles of models hosting a thermal inversion. The red profiles are the ones from the T/P grid described in Sect. \ref{subsec:the-model-spectra} and the black profiles are the MS models described in Sect. \ref{subsubsec:more-inversion-models}. The deep inversion layers of the MS models give rise to strong \ce{CO} emission.}
\label{fig:inversion-TP}
\end{figure}

For comparison, we calculated an additional suit of \ce{CO} models with a thermal inversion with T/P profiles qualitatively and quantitatively similar to the best-fit T/P profiles in \citet{Madhusudhan2009}. We call these the MS models. \citet{Madhusudhan2009} applied an atmospheric retrieval method with a parametric T/P profile to the Spitzer IRAC secondary-eclipse depths in \citet{Knutson2008}, supporting and constraining a dayside thermal inversion in HD 209458 b.  The T/P profiles of the MS models are shown in black in Fig. \ref{fig:inversion-TP} together with the profiles from the thermal inversion models of the T/P grid described in Sect. \ref{subsec:the-model-spectra} shown in red. We calculated the MS models using three different \ce{CO} volume-mixing ratios, $\textrm{VMR}=[\num{e-5}, \num{e-4}, \num{e-3}]$. 

The thermal inversions of the MS models are located deeper in the atmosphere than those from the T/P grid in the previous section. This results in strong emission in all cases and in line shapes that are very sensitive to the volume mixing ratio. The models with the lowest $\textrm{VMR}=\num{e-5}$ probe relatively low altitudes, making the inversion layer more spectroscopically dominant, thus resulting in line profiles with pure emission after convolution to CRIRES resolution. Models with higher VMRs still have siginificant emission due to the deep inversion layer, but they have absorption at the core of the lines because the core probes higher altitudes than for the lower VMRs where the temperature again decreases with increasing altitude. The shape of the lines at CRIRES resolution of MS models with a VMR of \num{e-4} or \num{e-3} resemble the line profile of the weak-inversion model in the middle panel of Fig. \ref{fig:model-spectra} at CRIRES resolution, only mirrored vertically. 

Note that recent analyses of the dayside spectrum of HD209458b by \citet{Diamond-Lowe2014} and \citet{Zellem2014} found that the data were best fit by a near-isothermal T/P profile. However, with high-dispersion spectroscopy we are only sensitive to the narrow components of the spectra, which are produced relatively high up in the atmosphere, making our observations insensitive to such a near-isothermal atmosphere. This is further discussed in Sect. \ref{sec:discussion}.

\subsection{Cross-correlation analysis}
\label{subsec:cross-correlation-analysis}

Each model was convolved to the CRIRES resolution using a Gaussian filter, then Doppler-shifted over the range \num{-250} to \SI{250}{\kilo\m\per\s} in steps of \SI{1.5}{\kilo\m\per\s}, and for each velocity step cross-correlated with the residual spectra. The velocity range was chosen to cover every possible radial velocity of the planet, and the step size roughly corresponds to the pixel velocity scale of the detectors. The outcome is a geocentric correlation matrix, with the strength of the cross-correlation as a function of the applied radial velocity shift and the spectrum number (i.e. time). The cross-correlation was performed separately for each night and detector. 

To maximise a potential planet signal, each row of the geocentric cross-correlation matrices, that is, the cross-correlation functions for each spectrum, was shifted to the rest frame of the planet, then summed over time. Finally, the summed, planetocentric cross-correlation functions from detectors 2 and 3 from all three nights were combined, providing the total cross-correlation functions (T-CCF).

The radial velocity corrections $(rv_\textrm{corr})$ from the geocentric to the planetcentric frame are a function of time and are given by
\begin{equation}
 rv_\textrm{corr}(t) = rv_\textrm{planet}(t) + rv_\textrm{helio}(t) + V_\textrm{sys}. 
\label{eq:rv-corr}
\end{equation}
Here, $rv_\textrm{planet}$ is the radial velocity curve of the planet, $rv_\textrm{helio}$ is the heliocentric correction term, and $V_\textrm{sys}$ is the systemic velocity of the host star \citep[-\SI{14.69 \pm 0.09}{\kilo\m\per\s},][]{Nidever2002}. Assuming a circular orbit \citep{Crossfield2012}, the radial velocity curve is given by
\begin{eqnarray} 
        rv_\textrm{planet}(t) = K_\textrm{P} \sin(2\pi \varphi(t)),
        \label{eq:rv-planet}
\end{eqnarray}
with $K_\textrm{P}$ being the semi-amplitude of the orbital radial velocity of the planet and $\varphi$ being the orbital phase. The orbital period and the mid-transit time used to calculate the phases of the planet are adopted from \citet{Knutson2007a}.

Since HD 209458 b is a transiting planet with an inclination of $i = 86.589^{\circ} \pm 0.076^{\circ}$ \citep{Southworth2008} close to $90$ degrees, $K_\textrm{P}$ equals the orbital velocity $V_\textrm{orb} = \SI{140 \pm 10}{\kilo\m\per\s}$ \citep{Snellen2010} to within the uncertainty. However, we investigated a wider range of $K_\textrm{P}$ values (50 -- 170 \si{\kilo\m\per\s}) and $V_\textrm{sys}$ values (-75 -- 45 \si{\kilo\m\per\s}) to scan the parameter space and check the strength of potential spurious signals.

\section{Results}
\label{sec:results}

\begin{figure*}[ht]
\center
\includegraphics[width=17cm]{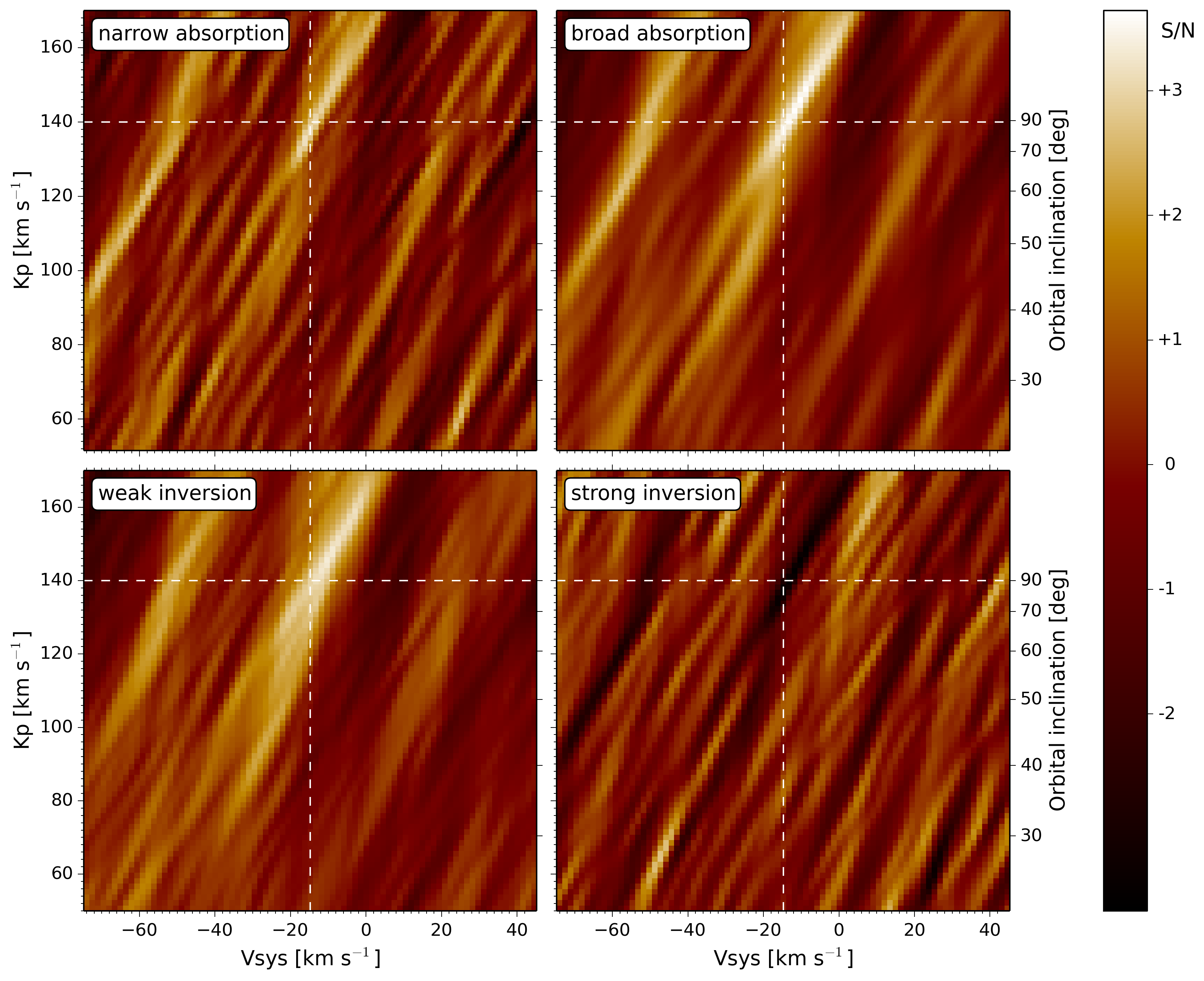}
\caption{Total cross-correlation diagrams for the four models illustrated in Fig. \ref{fig:model-spectra} after summing over time for a range of $V_\textrm{sys}$ and $K_\textrm{P}$ values, and combining detectors 2 and 3 from the three observation nights. The dashed white lines indicate the expected planet signal based on literature values, which coincides with a weak correlation signal in the case of the two absorption models as well as the weak-inversion model, and a weak anti-correlation signal in the case of the strong-inversion model. Although there are hints of a signal visible, \ce{CO} is not detected with statistical significance.}
\label{fig:finaldiag}
\end{figure*}

The diagrams in Fig. \ref{fig:finaldiag} show the strength of the total cross-correlation as function of $K_\textrm{P}$ and $V_\textrm{sys}$ for the four example models from Fig. \ref{fig:model-spectra}. The colour scale indicates the signal of the total cross-correlation function for a given parameter set. The S/N was obtained by dividing the total cross-correlation values by their standard deviation excluding points in parameter space close to the expected signal position. The $K_\textrm{P}$ values translate into orbital inclinations, which are given on the right-hand axis of the figures. The literature $K_\textrm{P}$ and $V_\textrm{sys}$ values are indicated by the dashed white lines. 

The absorption models and the weak-inversion models in the grid show a S/N = 3 -- 3.5 hint of a \ce{CO} correlation signal at a position within the uncertainties of the literature values of $K_\textrm{P}$ and $V_\textrm{sys}$. The models from the T/P grid with the strongest emission lines and only low-level absorption in the wings, such as shown in the bottom panel of Fig. \ref{fig:model-spectra}, give the lowest cross-correlation values. In fact, these models give rise to anti-correlation signals with S/N values of -2.5 -- -3.5. This is expected because the strong-inversion models are close to the inverse of the non-inversion models. The peak of the cross-correlation signal in the T-CCF diagram is in all cases very close (within 1 $\sigma$) to the expected position of the planet signal. The values of the peak cross-correlation signals, their positions, and the cross-correlation signal at the expected planet position for the four example models are all given in Table \ref{table:correlation-results}.

The $\sim3\sigma$ signal we obtain for non-inverted models is in our opinion not strong enough to claim an unambiguous detection of CO. In the T-CCF diagrams (Fig. \ref{fig:finaldiag}) one can see spurious signals that are only slightly lower in significance. Indeed, in previous analyses where we claimed planet signals, we obtained significantly higher S/N in the range S/N 4 -- 6 \citep{Snellen2010,Brogi2012,Birkby2013,deKok2013,Brogi2014}.

\begin{table}[h]
\caption{Results from the cross-correlation analysis for the four example models from Fig. \ref{fig:finaldiag}. The first column gives the S/N of the correlation or anti-correlation signal at the expected planet postion (-14.69, 140)\si{\kilo\m\per\s}. The second column gives the peak (anti-)correlation value, and the third column gives the location of the peak relative to the expected position. Note that these positions are all well within the 1$\sigma$ uncertainty.}             
\centering       
\begin{tabular}{r c c c}                
\hline\hline                            
\\
            & S/N  & S/N & peak location \\
            & at planet pos.  & peak        &  [km s$^{-1}$]\\
\hline                            
\\
narrow abs. &  2.8 &  3.2 & (-1.5, +4.5) \\
broad abs.  &  3.2 &  3.6 & (+4.5, +7.5) \\
weak inv.   &  3.0 &  3.4 & (+4.5, +7.5) \\
strong inv. & -2.8 & -3.6 & (+4.5, +7.5)\\
\hline                            
\end{tabular}
\label{table:correlation-results}
\end{table}

The diagonal noise structures in the T-CCF diagrams of Fig. \ref{fig:finaldiag} are a result of a degeneracy between $K_\textrm{P}$ and $V_\textrm{sys}$, with the slope depending on the observed orbital phases of the planet. The combination of three nights with slightly different phases, all with the planet on the same side of the host star, has acted to broaden and distort these naturally occurring noise structures. The noise structures are also affected by the line shape of the model. Cross-correlation with the narrow absorption model results in more narrow noise structures, and the location of the potential \ce{CO} signal is better defined. At the opposite end of the scale is the weak-inversion model, for which the line shapes at the CRIRES resolution resemble absorption doublets (see Fig. \ref{fig:model-spectra}). Cross-correlation with this much broader line profile smoothes the potential signal along the diagonal noise structures, increasing the uncertainty in the peak position. The peak cross-correlation values as given in Table \ref{table:correlation-results} all fall along the degenerate diagonal of the diagrams.

\subsection{MS models}
\label{subsec:MS-models}

The models with temperature-pressure profiles similar to the best-fit models of \citet{Madhusudhan2009} have two basic types of \ce{CO} line shapes at the CRIRES resolution, as described in Sect. \ref{subsubsec:more-inversion-models}. The pure emission lines can be seen as a mirror of the pure absorption lines from the T/P grid, and the lines with absorption at the core and strong emission in the wings are mirrors of the line shapes of weak-inversion models from the T/P grid. We do not detect carbon monoxide when cross-correlating the MS models with the observed residual spectra, as described in Sect. \ref{subsec:cross-correlation-analysis}. The S/N at the expected planet position is about $-3.3$ for the models with strong simple emission lines and about $-2.8$ for the models with some absorption at the core of the otherwise emission-dominated lines. The weak anti-correlations are consistent with the results from the T/P grid models, and indeed the T-CCF diagrams of the MS models look like colour-inverted versions of the diagrams for the models with broad absorption lines and weak inversion in Fig. \ref{fig:finaldiag}.

\subsection{Water models}
\label{subsec:water-models}

We have also cross correlated model spectra with water as the single trace gas. The water models were calculated using the same T/P grid as for the \ce{CO} models, but the tested volume-mixing ratios for the \ce{H2O} models were $\textrm{VMR}=[\num{e-5}, \num{e-4}, \num{e-3}]$. We did not detect a water signal from cross-correlating with any of the \ce{H2O} models. The typical correlation S/N at the expected planet position lies within $\pm 1$, and there is no systematic behaviour in the T-CCF diagrams as a function of the line profiles, such as for the \ce{CO} models. Recently, \citet{Madhusudhan2014} constrained the \ce{H2O} abundance of HD 209458 b to sub-solar levels assuming a clear atmosphere, with the central \ce{H2O} VMR measured to be \num{5.4e-6}, which is lower than the VMRs of the models in this work. The water abundance may still be solar or supersolar if the atmosphere of HD 209458 b has clouds or hazes.

\begin{figure*}[ht]
\centering
\includegraphics[width=17cm]{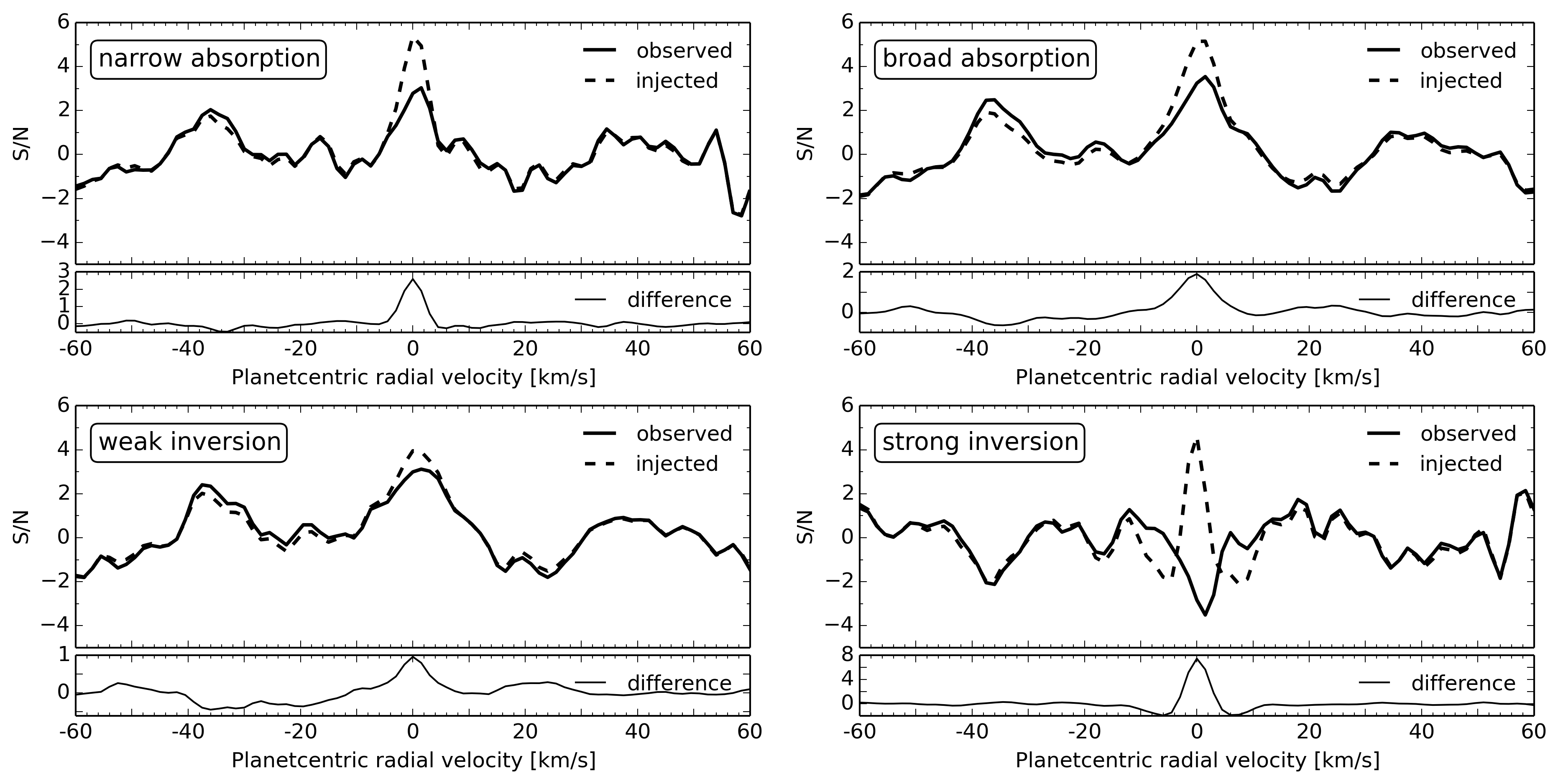}
\caption{Slices through the total cross-correlation diagrams for the four atmospheric models from Fig. \ref{fig:model-spectra}, assuming the literature values $V_\textrm{sys}=\SI{-14.69}{\kilo\m\per\s}$ and $K_\textrm{P}=\SI{140}{\kilo\m\per\s}$. The solid line is the T-CCF from cross-correlating a given model with the observed spectra, and the dashed line is the T-CCF from injecting the model at nominal strength into the observed spectra before telluric removal, and then cross-correlating with the same model. The differences between the observed and the injected T-CCFs are shown in the lower panels, and the peak signal in the difference T-CCFs are good estimates of expected S/N correlation signals, if the models are correct.}
\label{fig:1Dresult}
\end{figure*}

\subsection{Estimating expected CO signals}

We injected model spectra into the observed spectra as fake planet signals and subsequently retrieved the signals to evaluate the expected \ce{CO} signals as function of volume-mixing ratio and T/P profile.

First the model planet spectrum ($F_{\textrm{model}}$) was scaled with respect to the stellar continuum of the parent star. We adopted the effective temperature for HD 209458 of \SI{6092}{\K} \citep{Boyajian2015}, and approximated the stellar spectrum from a linear interpolation between two Castelli-Kurucz stellar atmosphere models\footnote{ftp://ftp.stsci.edu/cdbs/grid/ck04models/} with [M/H]=0.0, log g = 4.5 and the temperatures 6000 and 6250. The scaling was subsequently performed according to
\begin{eqnarray} 
        F_{\textrm{scaled}}(\lambda) = \frac{F_{\textrm{model}}(\lambda)}{F_{\textrm{S}}(\lambda)} 
        \left( \frac{r_{\textrm{P}}}{r_{\textrm{S}}} \right)^2,
\label{eq:scaling}
\end{eqnarray}
where $F_{\textrm{S}}$ is the model stellar spectrum, and $r_{\textrm{P}}$ and $r_{\textrm{S}}$ are the radii of the planet and star. The scaled model planet spectrum was then convolved to match the CRIRES resolution. For each observed spectrum the relative radial velocity between HD 209458 b and the observer was calculated using Eq. (\ref{eq:rv-corr}), and the scaled model planet spectrum was Doppler-shifted accordingly before being injected into the aligned and normalised observed spectra (step 2 in Fig. \ref{fig:waterfall}) using
\begin{eqnarray}
  F_{\textrm{obs+fake}}(\lambda) = F_{\textrm{obs}}(\lambda) \cdot \left( 1 + F_{\textrm{scaled}}(t, \, \lambda) \right).
\label{eq:inject-fake}
\end{eqnarray}
We then proceeded with the telluric removal as described in Sect. \ref{subsec:removal-of-telluric-contamination}, and the fake planetary signal was retrieved by cross-correlating with the same input model spectrum as described in Sect. \ref{subsec:cross-correlation-analysis}. 

Since the parameter $K_\textrm{P}=\SI{140}{\kilo\m\per\s}$ of the injected signal is precisely known, it is sufficient to examine slices through the T-CCF diagram when estimating the expected signals. Such one dimensional T-CCfs are shown in Fig. \ref{fig:1Dresult} for each of the four example models from Fig. \ref{fig:model-spectra}. The T-CCFs obtained from cross-correlating with the observed spectra are the solid lines, and the dashed lines are the T-CCFs from cross-correlating with the spectra injected with artificial planet signals. The injected planetary signatures show up as central peaks. We note that the shape of the four injected T-CCFs in Fig. \ref{fig:1Dresult} are to a high degree determined by the auto-correlation function of the average line profile of the model. This is especially true for the relatively large negative wings on either side of the correlation peak of the strong-inversion model. 

Since the injected signal is superimposed on top of the observed T-CCF, we need to examine the difference between the T-CCF with and without the injected signal, which are shown in a sub-panel below the T-CCFs in Fig. \ref{fig:1Dresult}. This shows that we can expect signals with a S/N of 2.6 for the narrow-absorption model, 1.9 for the broad-absorption model, 1.0 for the weak-inversion model, and 7.4 for the strong-inversion model. These results are to be interpreted as the approximate strength of the signal we should be able to detect if that spectral signature is present in the planetary atmosphere. 

Both of the absorption models have expected signals that are similar in strength to the observed signals, but because there are spurious signals of similar strength, more data are needed to confidently confirm the detection of \ce{CO} absorption lines. For the strong-inversion model, it is immediately clear that the observed correlation results are a poor match to the expected results. The emission lines are expected to give rise to a strong planetary signature, but there is no hint of this in the observed T-CCF. Instead, the observed T-CCF shows a weak anti-correlation signal, consistent with the possible presence of absorption lines. The weak-inversion model has an expected S/N $<$ 1, revealing that it is very difficult to detect models with such line profiles.

\subsubsection{Line contrast}

For the simplest line profiles, those with either pure absorption or pure emission, it is possible to place an upper limit on the line contrast, that is, the line strength relative to the stellar continuum. We performed this analysis for the two absorption models in the top panels of Fig. \ref{fig:model-spectra}. The model was injected into the observed spectra with variable strength by multiplying $F_{\textrm{scaled}}$ in Eq. (\ref{eq:scaling}) and (\ref{eq:inject-fake}) with a variable scaling factor $S$. For each value of $S$ the difference between the observed and injected T-CCFs was determined, and we measured the line contrast of the model $S*F_{\textrm{scaled}}(\lambda),$ which gave rise to a correlation signal with $\textrm{S/N} = 3$. This resulted in a line contrast of $\sim \num{3.2e-4}$ for both the narrow- and broad-absorption model.

\section{Discussion}
\label{sec:discussion}

We searched for a \ce{CO} signal in dayside spectra of HD 209458 b by cross-correlating with template spectra with \ce{CO} as the single trace gas, and assuming many different temperature-pressure profiles that covered atmospheres with and without a temperature inversion. The presence of an inversion layer gives rise to \ce{CO} emission in the spectra of the planet, but depending on the pressure range and temperature gradient of the inverted layer, the shape of the \ce{CO} lines can differ significantly. A thermal inversion will give rise to strong emission when it is located deep in the atmosphere and has a large temperature difference. The depth of a temperature inversion depends on the opacity sources responsible for the inversion, but since the sources are unknown in hot Jupiters, it is relevant to test models with inversions at different depths. At the high spectral resolution of the CRIRES instrument, more complex line shapes can be partly resolved, with the core of the lines probing higher altitudes, while the wings of the lines probe deeper, just above where the atmosphere becomes optically thick. The line shape for a given T/P profile is to a lesser degree also affected by the \ce{CO} volume-mixing ratio because the VMR influences the probed pressure range. 

We have tested models with a wide variety of line shapes: Pure absorption and pure emission lines, as well as models with either emission at the core and absorption in the wings, or absorption at the core and emission in the wings. We did not detect carbon monoxide in the HD 209458 b dayside spectrum with statistical significance, regardless of which model we cross-correlated with. The models with non-inverted atmospheres and the models with a weak-inversion layer, typically located at pressures lower than 0.01 bar, all show a hint of a correlation signal with $\textrm{S/N} = 3.0$ -- $3.5$, while the models with stronger inversion layers, including the MS models with T/P profiles like those of \citet{Madhusudhan2009}, show anti-correlation signals with similar low-significance levels. 

\begin{figure}
\centering
\includegraphics[width=8.5cm]{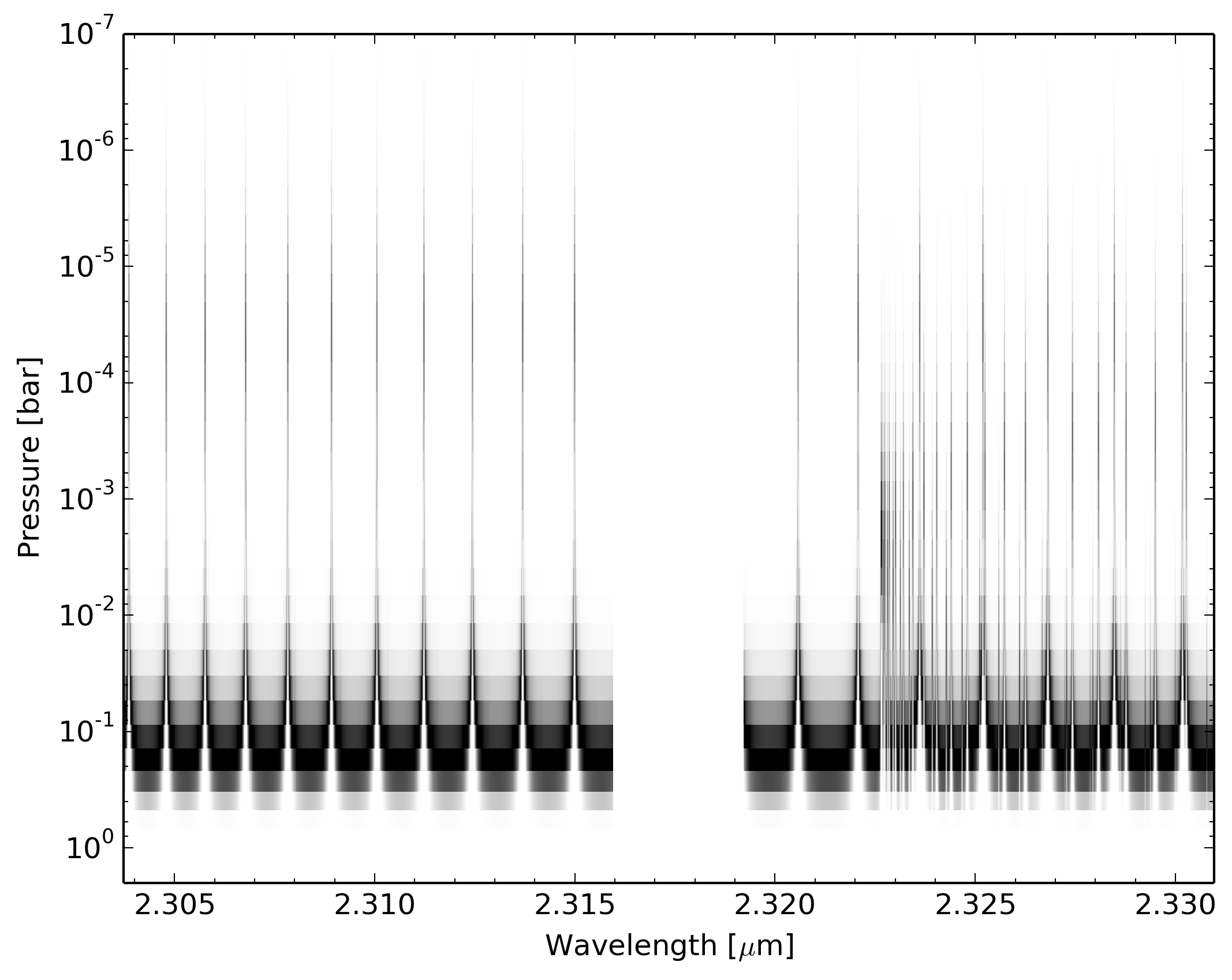}
\caption{Contribution function of the CRIRES observations for a CO volume-mixing ratio of 10$^{-3}$, with CO as a single trace gas. Note that these observations are sensitive to the narrow-line components over a pressure range $\sim$10$^{-2}$ to $\sim$10$^{-5}$ bar. In particular, in this range the T/P profile as derived by \citet{Diamond-Lowe2014} is near isothermal, making these observations insensitive to it.}
\label{fig:contribution-func}
\end{figure}

Our results are inconsistent with the inference of \citet{Knutson2008} of a thermal inversion in HD 209458 b based on Spitzer IRAC broadband photometry, but agrees with the more recent IRAC data and analysis by \citet{Zellem2014} and \citet{Diamond-Lowe2014}, who both found that a thermal inversion is not required to explain the Spitzer secondary-eclipse depths. In fact, the best-fitting T/P profile found by \citet{Diamond-Lowe2014} is isothermal at pressures lower than 0.1 bar. However, the high-resolution spectroscopy method is only sensitive to narrow components in the planetary spectrum - any broad component is filtered out in the early stages of the analysis. In Fig. \ref{fig:contribution-func} we show the contribution function for a CO volume-mixing ratio of 10$^{-3}$, indicating that our data are mostly sensitive to temperature differences in the 10$^{-5}$ to 10$^{-2}$ pressure range. This
means that we expect no signal from an atmosphere such as shown by \citet{Diamond-Lowe2014}. Our non-detection is therefore consistent with their results.

The non-detection of \ce{CO} in the dayside spectrum of HD 209458 b is interesting in light of previous successes with detecting \ce{CO} using high-resolution spectroscopy at \SI{2.3}{\micro\m}, both in the dayside spectra of other hot Jupiters \citep{Brogi2012, Rodler2012, Brogi2013, deKok2013, Brogi2014} and in the transmission spectrum of HD 209458 b itself \citep{Snellen2010}. \citeauthor{Snellen2010} detected \ce{CO} absorption and determined the line strength relative to the stellar continuum to be 1 -- \num{1.5e-3}, while in this work we obtain a best estimate of the relative line strength for simple absorption lines in the dayside spectrum to be \num{\sim 3.2e-4}. 

The most likely explanation for the lack of \ce{CO} signal is that the average dayside T/P profile is near-isothermal in the pressure range we probe (\num{e-1} -- \num{e-5} bar, see Fig. \ref{fig:contribution-func}). However, we note that at high altitudes there may still be a weak inversion layer, for instance at pressures $<<$ \num{e-3} bar, which are only probed by very narrow cores of the CO lines. Furthermore, when observing the dayside spectrum of HD 209458 b, the spectrum is disk integrated. HD 209458 b is expected to be tidally locked to its parent star, and \citet{Zellem2014} have observed the planet to have a hotspot, shifted eastward of the substellar point due to equatorial superrotation, as  expected if the planet is in synchronous rotation. It is possible that the planetary atmosphere has a thermal inversion only in the substellar point, or alternatively, in the eastward-shifted hotspot. If a thermal inversion is localised to only a part of the dayside, the effect of the inversion will be muted. 

Although high, the limited spectral resolution ($R=\num{100000}$) of the CRIRES instrument can work to dampen a thermal inversion signal somewhat. This is especially true for the weakest inversions (high altitude and less steep temperature gradient dT/dP), where the emission at the core of the lines is so weak that the complex line profiles are reduced to slightly muted, simple absorption lines. This effect alone cannot explain the lack of a \ce{CO} detection, but it may be a contributing factor.

Alternatively, the \ce{CO} signal, whether it is absorption or emission, might be heavily dampened by clouds or hazes. \citet{Charbonneau2002} first suggested an opaque cloud layer as a possible explanation for the lack of a strong optical Na feature in the optical transmission spectrum of HD 209458 b, although they noted that if clouds were the sole explanation, the cloud tops were required to be located above pressures of 0.4 mbar. It is unclear if such high-altitude cloud layers can exist in hot Jupiters, although \citet{Fortney2003} suggests it is possible to have clouds at pressures of several millibars.  However, one does expect that the contributions from clouds or hazes will be significantly stronger in transit spectra than in the dayside spectrum because of the grazing geometry in the former. \citet{Deming2013} and \citet{Madhusudhan2014} have observed the transmission spectrum of HD 209458 b with HST/WFC3 and both found weak transit features that possibly require haze. The haze alone is not expected to have a major influence on the thermal emission spectrum, but a cloud or haze layer at altitudes below those probed by transit observations might influence the continuum level for the thermal emission spectra and hence our results. The role of clouds or haze is thus still unclear.

\section{Conclusion}
\label{sec:conclusion}

We did not detect carbon monoxide at a statistically significant level in the \SI{2.3}{\micro\m} high-resolution dayside spectrum of HD 209458 b, although we saw a potential absorption signal at the 3 -- 3.5$\sigma$ level. \ce{CO} is expected to be abundant in hot Jupiters according to theory \citep[e.g.][]{Madhusudhan2012}, and has been found to be present in the tranmission spectrum of the planet \citep{Snellen2010}. A near-isothermal pressure profile of the planetary atmosphere is the most probable cause of the lack of CO signal. 

We showed that high-resolution spectroscopy can in principle measure line shapes and constrain the temperature pressure profiles in exoplanet atmospheres. 
In general, it will be possible to distinguish between a clear atmosphere with and without a thermal inversion, as long as these occur within the probed pressure range.
 
Our observations support the recent findings by  \citet{Zellem2014} and \citet{Diamond-Lowe2014} that there is no strong inversion layer in the dayside atmosphere of HD209458b, in the pressure range \SI{e-1}{\bar} -- \SI{e-3}{\bar}. The emission signal from such an inversion layer would have been readily detected in our spectra.

\begin{acknowledgements}
We are thankful to the ESO staff of Paranal Observatory for their support during the observations, and we thank the anonymous referee for the insightful comments and suggestions. This work is part of the research programmes PEPSci and VICI 639.043.107, which are financed by the Netherlands Organisation for Scientific Research (NWO). Support for this work was provided in part by NASA, through Hubble Fellowship grant HST-HF2-51336 awarded by the Space Telescope Science Institute. This work was performed in part under contract with the California Institute of Technology (Caltech)/Propulsion Laboratory (JPL) funded by NASA through the Sagan Fellowship Program executed by the NASA Exoplanet Science Institute.
\end{acknowledgements}


\bibliographystyle{aa}
\bibliography{../../mybib}

\end{document}